\documentclass[conference]{IEEEtran}
\IEEEoverridecommandlockouts

\usepackage[utf8]{inputenc}
\usepackage[T1]{fontenc}
\usepackage{graphicx}
\usepackage{amsmath}
\usepackage{amsthm}
\usepackage{amssymb}
\usepackage{units}
\usepackage{ifthen}
\usepackage{listings}
\usepackage{paralist}
\usepackage{xcolor}
\usepackage{xspace}
\usepackage{wrapfig}
\usepackage{float}
\usepackage[inline]{enumitem}

\graphicspath{{figures/}}

\usepackage[unicode=true,
  hidelinks,
  pdfauthor={Rafal Graczyk, Marcus Voelp, and Paulo Esteves-Verissimo},
  pdftitle={ADD FINAL TITLE HERE},
  pdfsubject={}, 
  pdfkeywords={},
  colorlinks=true,
  allcolors=black,
  citecolor=blue
]{hyperref}

\usepackage{cleveref}

\newboolean{showcomments}
\setboolean{showcomments}{true}

\ifthenelse{\boolean{showcomments}}
           { \newcommand{\mynote}[3]{
               \fbox{\bfseries\sffamily\scriptsize#1}
                    {\small$\blacktriangleright$\textsf{\emph{\color{#3}{#2}}}$\blacktriangleleft$}}}
           { \newcommand{\mynote}[3]{}}

\definecolor{vrpink}{RGB}{255,0,127}
\definecolor{orange}{RGB}{225,128,0}
\definecolor{brown}{RGB}{225,128,128}
\definecolor{purple}{rgb}{0.54, 0.17, 0.89}
\definecolor{rggold}{RGB}{218,165,32}

\ifthenelse{\boolean{showcomments}}
           { 
           }{}

\definecolor{darkgreen}{RGB}{0,155,0}
\begin{document}

\title{
  EphemeriShield - defence against cyber-antisatellite weapons}

\author{
   \IEEEauthorblockN{
   	Rafal Graczyk\IEEEauthorrefmark{1},
   	Marcus Voelp\IEEEauthorrefmark{1}, 
   	Paulo Esteves-Verissimo\IEEEauthorrefmark{1}}
   \IEEEauthorblockA{
	     CritiX Lab - Critical and Extreme Security and Dependability\\
     \IEEEauthorrefmark{1} SnT - Interdisciplinary Centre for Security,
     Reliability and Trust, University of Luxembourg}
     Email: \url{<name>.<surname>@uni.lu}}

\maketitle

\thispagestyle{plain} \pagestyle{plain}





Satellites, are both crucial and, despite common misbelieve, very fragile parts our civilian and military critical infrastructure. While, many efforts are focused on securing ground and space segments, especially when national security or large businesses interests are affected, the small-sat, new-space revolution democratizes access to, and exploitation of the near earth orbits. This brings new players to the market, typically in the form of small to medium sized companies, offering new or more affordable services. Despite the necessity and inevitability of this process, it also opens potential new venues for targeted attacks against space-related infrastructure. Smaller organizations, with less established revenue models, have a natural incentive to cut-corners in search for cost optimization and shorter time-to-market \cite{Ida18}.
\newline
While there are many classical anti-satellite (ASAT) weapons using various kinds of effectors (kinetic, RF, laser), we recently observed a proof of concept for a further smart attack of this kind \cite{Pav19}, following a more cybernetic approach to attack the information sphere, rather than causing direct energy transfer in orbit. In most cases, satellite operators know very well for their own space assets, where they are localized and with what orbital parameters they fly (e.g., from GNSS receivers on board of their spacecrafts or from their own tracking and ranging facilities). What these operators typically can’t derive just by their own means of observation are the locations and orbital parameters of other objects, such as active and inactive satellites, but more importantly, fields of space debris on a potential collision course with the operator’s space assets \cite{Kel06,Manual19}.
Instead, they have to obtain information about such objects by querying two-line element (TLE) debris files provided by Celestrack or Space-Track or any other source. The consequences of orbital collisions don’t have to be discussed here, as are already widely known \cite{Kes10}. To avoid them, orbital conjunction assessment aims at foreseeing possible close encounters, threatening the well being of satellites, monitoring changes in the orbital situation, and engaging in collision avoidance maneuvers, to the degree that propulsion or attitude manipulation allows this \cite{Nic17}.
\newline
 Cyber-ASAT \cite{Pav19} describes a method for altering TLE debris to orchestrate fake alarms and unnecessary collision avoidance maneuvers, targeting satellites to exhaust their fuel or jeopardize their availability during collision avoidance. The spectrum of TLE attack possibilities (altering or spoofing) is wide:
\begin{enumerate}
  \item intentional, by the Space Surveillance and Tracking (SST) system operator,
    altering the TLE at the source
  \item intentional, by external groups, altering the TLE provided to satellite operator
    (hacking into SST user front-end, man-in-the-middle attacks)
  \item unintentional, by error 
\end{enumerate}
The consequences of such TLE spoofing attacks are diverse. They may include fake collision avoidance alarms, spaning from seemingly low critical but unnecessary maneuvers and propellant loss (shortening mission lifetime or degrading system performance) to potentially orchestrated activity, launching organized attack campaigns on some third party, by tricking several space assets operators into undertaking actions that increase the collision likelihood with that 3rd party’s space assets.
\newline 
In this work, we propose a countermeasure to the presented problem that include distributed solution, which will have no central authority responsible for storing and disseminating TLE information. Instead, each of the peers participating to the system, have full access to all of the records stored in the system, and distribute the data in a consensual manner, ensuring information replication at each peer node. This way, single point of failure syndromes of classic systems, which currently exist due to the direct ephemerids distribution mechanism, are removed.
Our proposed solution is to build data dissemination systems using permissioned, private ledgers where peers have strong and verifiable identities, which allow also for redundancy in SST data sourcing. Each partner, providing object localization data, can be held responsible (or at least be identified) for low quality information and excluded in the future.
\newline
In our proposed solution, object data (unique identifier and ephemerids) is stored in a blockchain. The chaincode (i.e. operations associated with a designed blockchain system) runs in the system and defines an assets (in this case, orbital elements of objects under tracking) and transactions (in this case, instructions on how to modify an asset, i.e. algorithm criteria to decide whether to update objects’ orbital elements or requests for conjunction analysis and warnings distributions).
\newline
All information injected to the system, has to pass sanity checks, validating the feasibility envelope of provided parameters by numerical propagation with respect to last inputs, or with respect to other ephemerids provided by other SST data sources. An example algorithm is outlined in
\Cref{fig:ephentry}. After the values are checked, sanitized and accepted by chaincode governing the data entry, a transaction is added to the blockchain and made visible for all participants in the system, which update their ephemerids catalog. In case of discrepancy or detection of non-feasible SST data entries, a warning is raised for manual intervention. This way, discrepancies reveal measurement errors, attempts of intentional information modification or significant orbital maneuvers without prior information.
\newline
Clearly at least three type of peer nodes in the system can be identified: 
\begin{itemize}
\item SST data providers, attempting to update the database with new orbital elements as soon as they are detected by physical measurements;
\item SST data users, who are interested in obtaining up to date orbital elements information on objects they control; and, 
\item a third type, which commonly could be the same as the second, looking for conjunction analysis results, which will raise alerts on future proximity events. 
\end{itemize}

\begin{figure}
  \begin{lstlisting}
    @{\bf on new ephemerides entry $\mathit{(object_i, epoch_{new}, OE_{new})}$:}@
      $(OE_{old}, epoch_{old})$ := $retrive\_last\_ephemeris(object_i)$
      $OE_{prop}$ := $\mathit{propagate}(OE_{old}, epoch_{old}, epoch_{new})$
      if $\vert\mathit{OE_{new} - OE_{prop}}\vert$ < $\epsilon$:
        $append\_ephemerides(object_i, epoch_{new}, OE_{new})$
      else:
        $raise\_warning(object_i, epoch_{new}, OE_{new})$
  \end{lstlisting}
  \caption{Orbital Elements entry check algorithm}
  \label{fig:ephentry}
\end{figure}

The third type requires significant data processing capabilities, as conjunction candidate objects need to be selected (whose number could be substantial, especially on Low-Earth Orbit (LEO)). The orbits of such candidates need to be propagated for some time in the future (depending on quality of used models, usually no longer than a few days ahead) to check for possible close proximity passes.
However, with today’s high data processing capabilities, this shall not pose any organizational and technical problems, in a large majority of ground segment facilities.
\newline
In order to keep the blockchain size reasonable, only recent system snapshots (i.e., the state of all the objects in catalog) and transactions  relating to them need to be kept. This is possible due to fact that ephemerids older than 3 days become useless for predicting the current position and velocity of an object on orbit, as numerical propagators, but especially popular SGP4, used with TLEs, cannot take into account all the perturbations affecting orbital movement of bodies, which can be significant on LEO. The resulting propagation errors would be too large.
\newline 
As all the operations on the surveyed objects catalog are immutable and traceable (thanks to the permissioned-ledger based blockchain) it is possible to assign financial value to the use of the system, as incentive for participants to provide correct and high quality data. 
Eventually, such an additional mechanism would lead to the quite desirable outcome, where SST data and analysis providers could have (at least to some extent) their costs covered by the users of their data.
Incentives embedded in proposed mechanism, along with the independence from central authority, redundancy in data sources and data processing may lead to a democratization of access to SST information, which further contributes to the stability of the system, providing trust without the need of central enforcement.

\bibliographystyle{IEEEtran}
\bibliography{paper}



\end{document}